\documentclass[preprint,12pt]{elsarticle}
\usepackage{epsfig}
\usepackage{amssymb}
\usepackage{slashed}
\usepackage{amsmath}
\usepackage{cancel}

\journal{Nuclear Physics A}

\newcommand{\ga}{\alpha}
\newcommand{\gb}{\beta}
\newcommand{\gc}{\gamma}

\newcommand{\gk}{\kappa}

\newcommand{\gs}{\sigma}

\newcommand{\go}{\omega}

\newcommand{\gO}{\Omega}

\newcommand{\gC}{\Gamma}
\newcommand{\hh}{\frac{1}{2}}

\newcommand{\dd} {{\rm d}}

\newcommand{\be}{\begin{equation}}
\newcommand{\ee}{\end{equation}}
\newcommand{\ba}{\begin{eqnarray}}
\newcommand{\ea}{\end{eqnarray}}

\begin{document}

\begin{frontmatter}

\title{Transport coefficients of the quark-gluon plasma in ultrarelativistic limit}

\author{Stefano Mattiello}
\address{Institut f\"ur Theoretische Physik, Universit\"at Gie{\ss}en, D-35390 Gie{\ss}en, Germany}

%\sloppy

\begin{abstract}
The calculation of the transport coefficients of the strong quark-gluon plasma
(sQGP), i.e. sheer viscosity $\eta$, heat conductivity $\gk$ and bulk viscosity
$\zeta$, in the first Chapman-Enskog approximation is presented. 
Their formulation in terms of two-fold integrals depending
on the particle interaction, known as relativistic omega integrals, is derived
and their evaluation in the ultrarelativistic limit is worked out assuming
a cross section independent of the relative and total momentum of two colliding particles.
We find a suppression of the bulk viscosity and a pronounced temperature
dependence for the sheer viscosity and heat conductivity.
However, at high temperatures, they scale with the
third and second power of the temperature respectively as expected.
Furthermore, we find that all results in this ultrarelativistic expansion are dominated by the leading order contribution.
\end{abstract}


\end{frontmatter}
%\maketitle
%\twocolumn

\section{Introduction}

Transport properties of the strongly interacting quark-gluon plasma (sQGP), in particular
the bulk and shear viscosities, which describe the hydrodynamic response of
the system to energy and momentum density fluctuations, have attracted
remarkable attention in the last years.
The experimental findings of the heavy-ion collisions at the Relativistic Ion
Collider (RHIC) have pointed to a small specific sheer viscosity, i.e. the ratio sheer viscosity to entropy density $\eta/s$, and led to the announcement of the discovery of the nearly    
perfect fluidity of the sQGP~\cite{Gyulassy:2004zy,Shuryak:2004cy}.
The key role of the viscous corrections has been also addressed already by the very
first  $v_2$ data for Pb-Pb collisions at the LHC~\cite{Aamodt:2010pa}.
Indeed, the shear viscosity plays a crucial role in the estimation of the
initial condition of the heavy ion collisions~\cite{Mattiello:2012npa}.

Additionally, the hybrid model VISHNU that couples viscous
hydrodynamics for the macroscopic expansion of the QGP to the hadron cascade
model for the microscopic evolution of the late hadronic stage is a very
promising tool for a more precise extraction of the QGP shear viscosity~\cite{Song:2012tv}.
Therefore, several studies on shear and bulk viscosity have been performed in
the last few years, for the confined~\cite{NoronhaHostler:2008ju,Demir:2008tr,NoronhaHostler:2012ug} as well as
for the deconfined phase~\cite{Mattiello:2009db,Chakraborty:2010fr,Bluhm:2010qf,Wiranata:2012br,Cremonini:2012ny}.
Additionally, also for the thermal conductivity models have been developed~\cite{Braby:2009dw,Kapusta:2012zb}. 
Therefore a dynamic consistent calculation of all transport coefficients, i.e. the shear viscosity $\eta$, the bulk viscosity $\zeta$ and the heat conductivity $\gk$ within a model describing the strong coupling properties of the QGP is mandatory.
Recently, we have performed an investigation of the viscosity
$\eta$ of the sQGP~\cite{Mattiello:2009db} in a dynamical way within kinetic 
theory in the ultrarelativistic limit using the virial expansion approach
introduced in Ref.~\cite{Mattiello:2009fk}. 
In this approach the choice of the interaction between the partons in
the deconfined phase plays a crucial role to reproduce the thermodynamic
properties of the QGP in comparison to lattice QCD.           
By using a generalized classical virial expansion we have calculated the corrections to a
single-particle partition function starting from an interaction potential whose parameters are fixed
by thermodynamical quantities. From the same potential one can evaluate the transport cross section and then
the sheer viscosity~\cite{Mattiello:2009db}.

We have found $\eta/s\approx 0.097$ which is very close to the theoretical lower bound.
Furthermore, for $T\leq 1.8 T_{\rm c}$  the ratio is in the range
of the experimental estimates $0.1-0.3$ extracted from RHIC
experiments.

In the present work we test this approach to the bulk viscosity and the heat conductivity. In particular, we focus on the ultrarelativistic expansion of
the transport coefficients.
We use a relativistic extension~\cite{vanLeeuwen1971323} of the Chapman-Enskog method~\cite{Enskog:1917th} developed
in~\cite{VanLeeuwen:1973ix} to calculate the values of the transport
coefficients for a gas with arbitrary particle interaction.
The specific form of the interaction is encoded in the transport cross section.
Very recently, this method was reviewed in Ref.~\cite{Wiranata:2012br} for the
shear viscosity only. The aim of the authors was to perform a quantitative
comparison between the results of shear viscosities from the Chapman-Enskog
and relaxation time methods for selected test cases with specified and
simplified elastic differential cross sections.
In this contribution we use the Chapman-Enskog method
for one specific quark-quark interaction. This leads to an equation of state in line with recent
three-flavor QCD lattice data for the
pressure, speed of sound, and interaction measure at nonzero
temperatures and vanishing chemical potential. Furthermore, from the same
interaction we have directly extracted the effective
coupling $\ga_{\rm V}$ to be employed for the determination of the transport
cross section which enters the transport coefficients. In this way, we systematically
calculate 
all three transport coefficients $\eta$, $\zeta$ and $\kappa$.

This paper is organized as follows.
In Section~\ref{Sec:Theory}, following Refs.~\cite{Wiranata:2012br,Kox:1976xvi} we review the formal expression for the
transport coefficients in the first order of the first Chapman-Enskog approximation.
In particular, we give a formulation in terms of relativistic omega integrals.
In the ultrarelativistic limit these integrals are evaluated for cross sections which are independent on the relative
and total momentum of two colliding particles.

In Section~\ref{Sec:Results} we calculate and discuss the transport
coefficients for our model within the ultrarelativistic approximation.
%and our realistic Equation of state for the QGP is used to determine the
%specific sheer viscosity.
The conclusions in Section~\ref{Sec:Concl} finalize this work.

\section{Theory}\label{Sec:Theory}

A elegant and clear derivation of the general expression for the transport
coefficient  in the first Chapman-Enskog approximation can be found in Ref.~\cite{VanLeeuwen:1973ix}.
In the following we directly recall the results for the heat conductivity and
the sheer and bulk viscosity in the first order of this approximation of a gas of free
particles with mass $m$ at temperature $T$
\ba\label{def:eta}
\eta&=&\frac{1}{10}T\frac{\gc^2_0}{c_{00}},\\
\label{def:kappa}
\gk&=&\frac{1}{3}\frac{T}{m}\frac{\gb^2_{1}}{b_{11}},\\
\label{def:zeta}
\zeta&=&T\frac{\ga^2_2}{a_{22}}.
\ea
The quantities $\alpha_2,\gb_1$ and $\gamma_0$ are known functions of the
temperature and independent of the particle interaction.
In general, by introducing the dimensionless quantity
\be
z=m/T,
\ee
the enthalpy per particle
\be
h=\frac{K_3(z)}{K_2(z)}
\ee
and the adiabatic coefficient
\be
\gc=\frac{c_{p}}{c_V}
\ee
these scalar function are given by
\ba\label{eqn:gc00}
\gc_0&=&-10 h=-10\frac{K_3(z)}{K_2(z)},\\
\label{eqn:gb1}
\gb_1&=&-3\frac{\gc}{\gc-1}=\frac{\dd \{K_3(z)/K_2(2)\}}{\dd z^{-1}}\\
\label{eqn:ga2}
\ga_2&=&\frac{3}{2}\left[zh\left(\gc-\frac{5}{3}\right)+\gc\right],
\ea
where $K_n(z)$ denotes the modified Bessel function of the second kind of order
$n$.
The quantities $a_{22}$, $b_{11}$ and $c_{00}$ are
interaction independent and formally given by
 \ba\label{eqn:c00}
c_{00}&=&16\left(\go^{(2)}_2-z^{-1}\go^{(2)}_1+\frac{1}{3}z^{-2}\go^{(2)}_0\right),\\\label{eqn:b11}
b_{11}&=&8\left(\go^{(2)}_1+z^{-1}\go^{(2)}_0\right),\\
\label{eqn:a22}
a_{22}&=&2\go^{(2)}_0,
\ea
where the relativistic omega integrals are in general defined as 
\ba
\go^{(s)}_i(z)&=&\frac{2\pi z^3}{K^2_2(z)}\int_0^\infty \!\!\!\!\dd u \;\sinh^7 u \;\cosh^i u\;K_j(2z\cosh u)\nonumber\\
&\times&\int_0^\infty \!\!\!\!\dd\theta\; \sin \theta
\;\frac{\dd\gs(u,\theta)}{\dd \gO}\;
(1-\cos^s\theta),\qquad j=\frac{5}{2}+\hh(-1)^i.
\label{def:omega-s-i}
\ea
The particle interaction is encoded in the differential cross section
$\dd\gs(u,\theta)/\dd \gO$.
Here $\theta$ denotes the scattering angle in the center of momentum frame.
The variable $u$ is directly connected to the relative momentum $q_{12}$ and
total momentum $Q$ of the two colliding particles by
\be
q_{12}=m\sinh u \qquad {\rm and}\qquad Q=2m\cosh u.
\ee 
One has to evaluate the omega integrals in order
to calculate the transport properties of the system.
The results, in particular, the behavior of the cross section as function of
$u$ and $\theta$, are of course model dependent.
The simplest case, i.e. constant cross section, which correspond to
an assumption of hard spheres, has been investigated in Refs.~\cite{Wiranata:2012br,Kox:1976xvi}.
In following we consider the more general case, where the cross section is momentum
independent, i.e. $\dd\gs/(\dd \gO)=\dd\gs(\theta)/\dd \gO$ and we evaluate the omega integrals under this assumption.

\subsection {Omega integrals in ultrarelativistic limit}\label{Subsec:Omega}

The momentum independence of the differential cross section leads to a
simplification in Eq.(\ref{def:omega-s-i}).
Performing the the integration the $u$ integration, using $x=\cosh u$, yields 
\be
\go^{(s)}_i(z)=\frac{2\pi z^3}{K^2_2(z)}W^{(s)}\int_1^\infty
\!\!\!\!\dd x \;(x^2-1)^3\;x^i\;K_j(2zx),%, \qquad j=\frac{5}{2}+\hh(-1)^i,
\ee
where the angular integral $W^{(s)}$ is given by
\be
W^{(s)}=\int_0^\infty \!\!\!\!\dd\theta\; \sin \theta
\;\frac{\dd\gs(\theta)}{\dd \gO}\;(1-\cos^s\theta).
\ee
We note here, that for the calculation of the transport coefficients omega integrals for $s=2$ have to be computed.
Then, one can directly connect the integral
$W^{(2)}$ with the (viscous)transport section defined by~\cite{Danielewicz:1984ww}
\be\label{def:cross}
\sigma_{\rm t}\equiv \int \!\!\dd\gO \;\frac{\dd\sigma}{\dd \gO}\;\sin^2\theta.
\ee 
One obtains the relation
\be
W^{(2)}=\frac{\sigma_{\rm t}}{2\pi}
\ee
and the relativistic omega integral may be written as
\be
\go^{(2)}_i(z)=\frac{\gs_{\rm t}}{K^2_2(z)}z^3\int_1^\infty
\!\!\!\!\dd x \;(x^2-1)^3\;x^i\;K_j(2zx).%, \qquad j=\frac{5}{2}+\hh(-1)^i.
\ee
By replacing $zx\rightarrow x$ and using the binomial formulae we obtain
\be\label{eqn:omega-fin}
\go^{(2)}_i(z)=\frac{\gs_{\rm t}}{K^2_2(z)}z^{-i-4}\sum_{k=0}^3\binom{3}{k}(-1)^k z^{2k}\int_z^\infty
\!\!\!\!\dd x \;x^{i+6-2k}\;K_j(2x).%, \qquad j=\frac{5}{2}+\hh(-1)^i.
\ee

Because the limit $z\rightarrow 0$
corresponds to the ultrarelativistic limit, an expansion of $\go^{(2)}_j$
around $z=0$  leads to a systematic evaluation of the transport coefficients in
this approximation.
A detailed investigation of this expansion allows to estimate
the importance of the leading order terms as well as to systematically compare
the different transport coefficients at the same order in $z$.
In fact, in the ultrarelativistic limit, the integral in
Eq.(\ref{eqn:omega-fin}) becomes independent of $z$, because the lower bound
can be replaced in this expression by zero.
This replacement allows to perform the integration analytically using
\be\label{eqn:K-int}
\int_0^\infty\!\!\!\!\dd x \;x^\mu\;K_\nu(2x)=\frac{1}{4}\gC[(\mu+\nu+1)/2]\gC[(\mu-\nu+1)/2]
\ee
for $\mu\pm\nu+1>0$ (see Appendix~\ref{Sec:AppK}).
for the calculation of the transport coefficients this condition is
satisfied. 
In the ultrarelativistic limit the omega integral may be written as 
\be\label{eqn:omega-UR}
\go^{(2)}_i(z)=\frac{\gs_{\rm t}}{4 K^2_2(z)\;z^{i+4}}\sum_{k=0}^3\binom{3}{k}(-1)^k z^{2k}\gC[(\mu+\nu+1)/2]\gC[(\mu-\nu+1)/2].%, \qquad j=\frac{5}{2}+\hh(-1)^i.
\ee
We emphasize here, that this result is a fractional expression, where
the numerator is given by a polynomial in $z$.
The denominator contains the modified Bessel function $K_2(z)$, whose expansion
at $z=0$ can not be expressed as a polynomial as shown in
Appendix~\ref{Sec:AppK}. For $K_2(z)$ the leading terms of the expansion are 
\be\label{eqn:K_2}
K_2(z)=\frac{2}{z^2}\left[1-\frac{1}{4}z^2+-\frac{1}{16}z^4\ln
  z+\frac{1}{32}c_{\rm EM}z^4+O(z^6\ln z)\right],
\ee
where the constant $c_{\rm EM}$ is connected to the gamma of Euler-Mascheroni
$\gamma_{\rm EM}$ by
\be
c_{\rm EM}=2\ln 2+\frac{3}{2} -2\gc_{\rm EM}\approx 1.731863.
\ee
We obtain for the omega integrals the the following expansion in $z$
\ba
\go^{(2)}_i(z)&=&\frac{\gs_{\rm t}}{16}\gC[(i+j+7)/2]\gC[(i-j+5)/2]\nonumber\\
&\times&\left(z^{-i}+\frac{i^2+j^2+10i+1}{i^2-j2+10i+25}\frac{z^{2-1}}{2}\right)
  +o(z^{4-i}\ln z),\label{eqn:omegaUR}
\ea
where only the first two leading terms are considered.
Combining this expression with the analytic expression for the quantities
$\alpha_2,\gb_1$ and $\gamma_0$ given in Eqs.(\ref{eqn:gc00}-\ref{eqn:ga2}) we
can analytically evaluate the transport coefficients.

\subsection {Evaluation of transport coefficients in ultrarelativistic
  limit}\label{Subsec:Coeff}

For the calculation of the transport coefficients in the ultrarelativistic
limit we also need the asymptotic expansion for the scalar function
$\alpha_2,\gb_1$ and $\gamma_0$.
Using Eqs.(\ref{eqn:gc00}-\ref{eqn:ga2}), they can be expressed in term of the Bessel function $K_2(z)$ and $K_3(x)$
In addition to the asymptotic expansion $K_2(z)$ given in
Eq.(\ref{eqn:K_2}), we need the corresponding expression for $K_3(z)$,
\be\label{eqn:K_3}
K_3(z)=\frac{8}{z^3}\left[1-\frac{1}{8}z^2+\frac{1}{64}z^4+O(z^6\ln z)\right].
\ee  
Then we obtain for the (relevant) scalar quantities
\ba\label{eqn:gc002-UR}
\gc^2_0&=&\frac{1600}{z^2}\left(1+\frac{1}{4}z^2\right)+O(z^2\ln z),\\
\label{eqn:gb12-UR}
\gb^2_1&=&144\left(1-\frac{1}{4}z^2\right)+O(z^4\ln z),\\
\label{eqn:ga22-UR}
\ga^2_2&=&\frac{1}{36}z^4+O(z^6\ln z).
\ea 
The dynamical quantities $a_{22}, b_{11}$ and $c_{00}$ defined in
Eqs.(\ref{eqn:c00})-(\ref{eqn:a22}) can be evaluated in the ultrarelativistic
limit using Eq.(\ref{eqn:omegaUR}). We obtain
\ba\label{eqn:c00-UR}
c_{00}&=&\frac{200\gs_{\rm t}}{z^2}\left(1+\frac{1}{5}z^2+O(z^4\ln z)\right),\\
\label{eqn:b11-UR}
b_{11}&=&\frac{288\gs_{\rm t}}{z}\left(1+O(z^4\ln z)\right),\\
\label{eqn:a22-UR}
a_{22}&=&\frac{3\gs_{\rm t}}{2}\left(1-\frac{1}{4}z^2+O(z^4\ln z)\right).
\ea
Finally, using Eqs.(\ref{def:eta})-(\ref{def:zeta}) we can write the transport coefficients as
\ba\label{eqn:eta-UR}
\eta&=&\frac{4m}{5\gs_{\rm t}z}\left(1+\frac{1}{20}z^2+O(z^4\ln z)\right),\\
\label{eqn:kappa-UR}
\gk&=&\frac{4}{3\gs_{\rm t}}\left(1-\frac{1}{4}z^2+O(z^4\ln z)\right),\\
\label{eqn:zeta-UR}
\zeta&=&\frac{mz^3}{108\gs_{\rm t}}\left(1+O(z^2\ln z)\right).
\ea
Using the definition of $z$ for the leading term of the shear
viscosity we recover the known expression~\cite{Mattiello:2009db,Danielewicz:1984ww}
\be
\eta=\frac{4T}{5\gs_{\rm t}}\left(1+\frac{1}{20}z^2+O(z^4\ln z)\right),
\ee
We note that the leading term is temperature dependent, but independent
of the mass. For the heat conductivity $\gk$ the leading term does not depend on
$z$.
Nevertheless, a implicit temperature dependence can be encoded via the transport
cross section.
Additionally, we note that in this ultrarelativistic approach the bulk viscosity is automatically
suppressed in comparison to the other transport coefficient as expected from
general physical considerations~\cite{Landau-Lifshitz:X}.

\section{Results}\label{Sec:Results}
For the final calculation using
Eqs.(\ref{eqn:eta-UR}-\ref{eqn:zeta-UR}) we have to specify an interaction.
We follow Ref.~\cite{Mattiello:2009db} to
calculate the $\gs_{\rm t}$ from the interaction $W_{12}$ used to describe
within a virial expansion the three-flavor thermodynamics properties of the QGP from
the lattice~\cite{Cheng:2007jq}.
We use an effective quark-quark
potential $W_{12}$ motivated in Ref.~\cite{Mattiello:2009fk} and inspired
by a phenomenological
model which includes non-perturbative effects from dimension two gluon
condensates~\cite{Megias:2007pq}.
The effective potential between the quarks reads
\begin{equation} \label{pot}
W_{12}(r,T)=\left(\frac{\pi}{12}\frac{1}{r}+\frac{{\mathcal C}_2}{2N_{\rm c}T}\right){\rm e}^{-M(T)r},
\end{equation}
where ${\mathcal C}_2=(0.9 {\rm GeV})^2$ is the non-perturbative dimension two condensate and $M(T)$ the
Debye mass estimated as
\begin{equation}\label{def:Debyemass}
M(T)=\sqrt{N_{\rm c}/3+N_{\rm f}/6}\; gT=\tilde g T.
\end{equation}
Setting the coupling parameter $\tilde g=1.30$ one describes the recent
three-flavor QCD lattice data for all thermodynamic
quantities~\cite{Cheng:2007jq} in the temperature range up to 5 $T_{\rm c}$
very well.

The transport cross section
is given by~\cite{Zhang:1999rs,Molnar:2001ux}
\be \label{cross}
\sigma_{\rm t}(\hat s) =\sigma_0 \;4\hat z(1+\hat z)\left[(2\hat
  z+1)\ln(1+1/\hat z) - 2\right],
\ee
with the total cross section
$\sigma_0(\hat s) = 9\pi\alpha_{\rm V}^2(\hat s)/2\mu_{\rm scr}^2$.
Here $\ga_{\rm V}=\ga_{\rm V}(T)$ is the effective
temperature-dependent coupling constant and  and
$\hat z\equiv M(T)^2/\hat s$. For simplicity, we
will assume $\sigma_0$ to be energy independent and neglect its
weak logarithmic dependence on $\hat s$ in the relevant energy
range and set $\hat{s}\approx 17 T^2$~\cite{Majumder:2007zh}.

In this study we consider only elastic cross sections. An extension to
inelastic scattering processes for an ultrarelativistic Boltzmann gas and its
application for the calculation of the shear viscosity has been recently presented~\cite{El:2012cr}.
Following Refs.~\cite{Mattiello:2009db,Kaczmarek:2005ui} we can extract the
coupling $\ga_{\rm V}$ from the interaction $W_{12}$.
We define the coupling in the
so-called ${\rm qq}$-scheme,
\begin{eqnarray}
\alpha_{\rm qq}(r,T)&\equiv&-\frac{12}{\pi}r^2\frac{{\rm d}W_{12}(r,T)}{{\rm d}r}\;.
\label{alp_qq}
\end{eqnarray}
\begin{figure}[h!]
\begin{center}
    \epsfig{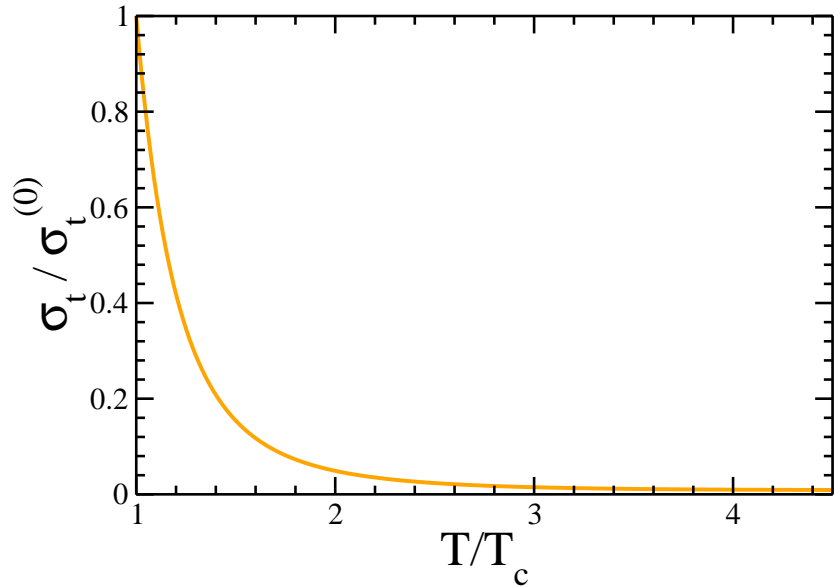}
\end{center}
\caption{\label{fig:Sigma_t-N}
(Color online) The transport cross section  normalized to its value at the
  critical temperature $\gs_{\rm t}/\gs_{\rm t}^(0)$
as a function of  
the temperature expressed in units of the critical temperature    
$T_{\rm c}$.}
\end{figure}

The coupling $\alpha_{\rm qq}(r,T)$ then exhibits a maximum for fixed temperature at a certain distance denoted by
$r_{\rm max}$. By analyzing  the size of the maximum at $r_{\rm max}$
 we fix the temperature dependent coupling $\alpha_{\rm V}(T)$ as

\begin{eqnarray}
\alpha_{\rm V}(T)&\equiv&\alpha_{\rm qq}(r_{\rm max},T)\;.\label{alp_Tdef}
\end{eqnarray}
With these results we obtain a temperature dependent transport cross section
$\gs_{\rm t}(T)$.

In Fig.~\ref{fig:Sigma_t-N} the
transport cross section normalized to its value at the critical temperature, is shown as a function of  
the temperature expressed in units of the critical temperature    
$T_{\rm c}$.
We note a strong decrease of the cross section with temperature. This can
be not neglected in a consistent evaluation of the transport
coefficients.

\begin{figure}[tbh!]
\begin{center}
    \epsfig{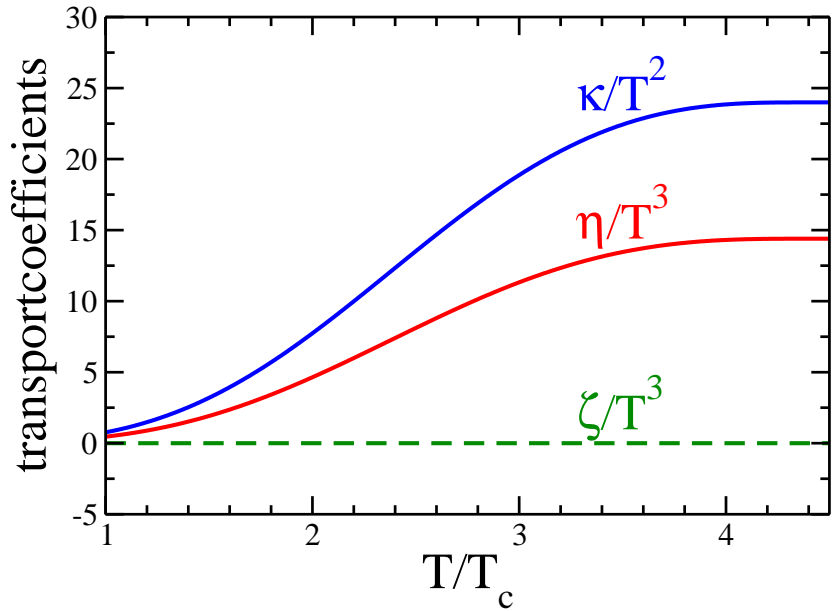}
\end{center}
\caption{\label{fig:Coeff-UR}
(Color online) The scaled sheer viscosity $\eta/T^3$ (red line), heat
  conductivity $\kappa/T^2$ (blue line) and bulk viscosity
$\zeta/T^3$ (green line) as function of  
the temperature expressed in units of the critical temperature    
$T_{\rm c}$.}
\end{figure}
\begin{figure}[tbh!]
\begin{center}
    \epsfig{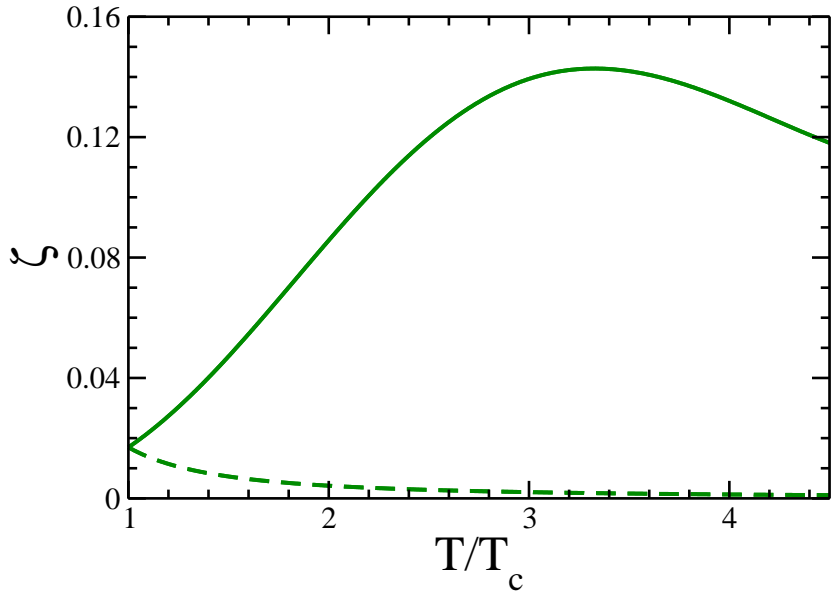}
\end{center}
\caption{\label{fig:zeta}
(Color online) The bulk viscosity $\zeta$ (solid line) as function of  
the temperature expressed in units of the critical temperature    
$T_{\rm c}$. In comparison the bulk viscosity using the temperature independent
transport cross section $\gs_{\rm
t}^{(0)}$.}
\end{figure}

In order to
investigate the importance of the different coefficients we scale them by the
appropriate power of the temperature in order to obtain a dimensionless
quantity.
Therefore we show in Fig.~\ref{fig:Coeff-UR} the ratios $\eta/T^3$ (red line),
$\kappa/T^2$ (blue line) and
$\zeta/T^3$ (green line) as a function of  
the temperature expressed in units of the critical temperature    
$T_{\rm c}$.
Several features become evident:
\begin{itemize}
\item[{\em i)}] For the shear viscosity and the heat conductivity an increase with the temperature is found
for $T_{\rm c}\lesssim T\lesssim 3.5T_{\rm c}$. 
\item[{\em ii)}] At higher temperatures $\eta/T^3$ and $\gk/T^2$ become constant. This behavior of the shear viscosity is also present in very different approaches, such as AdS/CFT~\cite{Gubser:1998nz}, the quasiparticle
approximation including the quark selfenergy~\cite{Alberico:2007fu,Czerski:2007ns}
and in the weak coupling estimate from Ref.~\cite{Thoma:1991em,Heiselberg:1994vy}.
For the heat conductivity that is also found in perturbative
calculations~\cite{Braby:2009dw,Heiselberg:1993cr}, where $\gk$ scales with the temperature square.
\item[{\em iii)}] The bulk viscosity is completely
  negligible in comparison to the other coefficients.
This justifies the omission of $\zeta$ in
several hydrodynamical and transport
calculations~\cite{Huovinen:2008te,Randrup:2010ax,Song:2010aq}.
\end{itemize}

In Fig.~\ref{fig:zeta} we show $\zeta$ as function of  
the temperature expressed in units of the critical temperature    
$T_{\rm c}$ (solid green line).
To investigate the role of the temperature dependent transport cross section
we compare with the results obtained using the constant cross section $\gs_{\rm
t}^{(0)}$ (green dashed line)
In the calculation with the full cross section the suspected enhancement of the bulk viscosity is
missing. In contrast, the interplay between cross section and $z^3$ leads to a
broad maximum around $3.2T_{\rm c}$.
This is clearly a consequence of the temperature dependence of $\gs_{\rm
t}$: namely, using the constant value $\gs_{\rm
t}^{(0)}$ for the transport cross section, we find a decreasing behavior, that suggests the possibility of a maximum of the bulk viscosity at $T_{\rm c}$.
It is not surprising that our ultrarelativistic model can not
describe the peak of $\zeta$ at the critical temperature.
This maximum is a consequence of the hadronic correlations
at $T_{\rm c}$~\cite{Paech:2006st} that are non included in this approach.
To include such correction one can add to this parameter free derivation of the
cross section an estimation of the hadronic transport cross section
%As first estimation we consider for the hadronic sector pions contribution
%only.
%Then the correlations term of the bulk viscosity in reads
%\be
%\zeta_{\rm corr}=\frac{M_\pi z_\pi^3}{108\gs_{\rm t}}\left(1+O(z^5\ln z)\right).
%\ee
%Following~\cite{Danielewicz:1984ww} we use for the hadronic transport cross
%section is 
%In order to implement that the limitation of the ultrarelativistic limit have to
%be relaxed 
and the omega integrals for the whole temperature range have to be
calculated. Therefore, analytic expressions for the coefficients can not
performed~\cite{Kox:1976xvi}. 

Alternatively, an estimate for the correlation term of the the bulk viscosity
can be given using the expression $\zeta_{\rm cor}=-A(c_{\rm
  s}^2-1/3)\eta$ , with $A=4.558$ or $A=4.935$~\cite{Benincasa:2005iv} or
$A=2$~\cite{Buchel:2005cv} respectively.

Another important aspect of our results regarding the sheer viscosity and the
heat conductivity is the relative importance of the next to leading order
contributions.
\begin{figure}[th!]
\begin{center}
    \epsfig{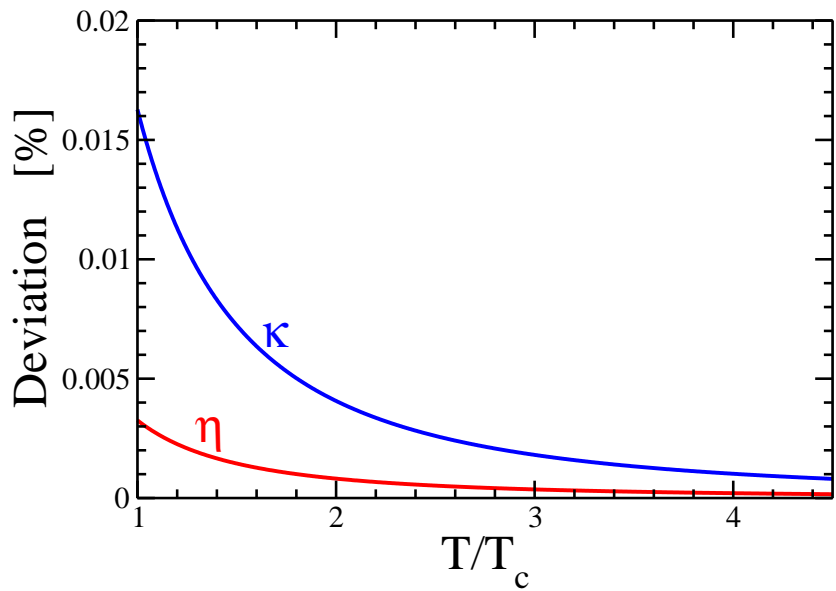}
\end{center}
\caption{\label{fig:Dev}
(Color online) The percentage deviation sheer viscosity $\Delta\eta$ (red
  line) and of the heat
  conductivity $\Delta\kappa$ (blue line)  as function of  
the temperature expressed in units of the critical temperature    
$T_{\rm c}$.}
\end{figure} 
In Fig.~\ref{fig:Dev} the percent deviation $\Delta\eta$ of the sheer
viscosity (red line) and $\Delta\gk$ of the heat conductivity (blue line) have been investigated.
The percent deviations are defined by
\be
\Delta X= \frac{\vert X-X^{({\rm NLO})}\vert}{X}\times 100,
\ee
where $X=\eta,\gk$ and  $X^{({\rm NLO})}$ denotes the next to
  leading order contribution.

Evidently, the deviations from the leading order terms are very small and the
higher order contributions can be neglected.
In this sense, the calculations of the ratio sheer viscosity to entropy
density performed in Ref.~\cite{Mattiello:2009db}, where the leading term of
the ultrarelativistic expansion and the entropy within this virial expansion
model have been used, is unchanged by the higher term corrections.

\section{Conclusions}\label{Sec:Concl}
We presented the calculation of the
transport coefficients of the QGP by revisiting the first order of the first Chapman-Enskog approximation.
We formulated them in terms of the relativistic omega integrals.
Their  evaluation in the ultrarelativistic limit is explictly shown by assuming
that the cross section is independent on the relative
and total momentum of two colliding particles.
The well known features of the ultrarelativistic limit are recovered,
i.e. the suppression of the bulk viscosity in comparison to the other coefficients. 
In particular, the expected peak of $\zeta$ at the critical temperature is missing,
because the hadronic correlations at $T_{\rm c}$ have not been included.
However, the overall suppression justifies neglecting of the bulk viscosity in
several calculations for the evolution of the QGP~\cite{Mattiello:2012npa,Huovinen:2008te,Randrup:2010ax,Song:2010aq}.
For the sheer viscosity and heat conductivity we observe an increasing behavior
with the temperature.
At high temperatures, they scale as power law in $T$.
Additionally, we find that all results in this ultrarelativistic expansion are
dominated by the leading order contribution. 
Therfore, for the ratio sheer viscosity to entropy density we recover the
results of our previous estimation~\cite{Mattiello:2009db}.

For further work, the limitation of the ultrarelativistic limit has to be
relaxed and the coefficients can be evaluated in the whole temperature range.
This allows the including of hadronic correlations at $T_{\rm
  c}$ in the calculations of $\zeta$.
Additionally, we can implement the temperature dependent transport coefficients calculated here as well as our realistic equation of state in dissipative hydrodynamical calculation.  
In this way we can systematically investigate the
evolution of the QGP in the heavy-ion collisions following the lines of Ref.~\cite{Mattiello:2012npa}.

Acknowledgment: 
I thank Stefan Strauss and Hendrik van
Hees for useful discussions and suggestions.
This work is supported by Deutsche Forschungsgemeinschaft and by the Helmholtz International Center for FAIR  
within the LOEWE program of the State of Hesse.

\appendix

\section{Modified Bessel functions of second kind}\label{Sec:AppK}

The Bessel functions~\cite{Abramowitz}, first defined by the mathematician Daniel Bernoulli and
generalized by Friedrich Bessel, are canonical solutions $w(z)$ of Bessel's
differential equation
\be
z^2 \frac{\dd^2 w}{\dd z^2} + x \frac{\dd w}{\dd z} + (z^2 - \nu^2)w = 0
\ee 
for an arbitrary real or complex number $\nu$ which indicate the order of the
Bessel function.
In general, $\nu$ is integer or half-integer.
The solutions called Bessel functions of the first kind are denoted as
$J_\nu(z)$ and are holomorphic functions of $z$ throughout the $z$-plane cut
along the negative real axis.
It is possible to define the function by its Taylor series expansion around $z = 0$
\be\label{eqn:J-exp}
J_\nu(x) = \frac{1}{2}z\sum_{k=0}^\infty \frac{(-\frac{1}{4}z^2)^k}{k! \, \Gamma(\nu+k+1)},
\ee where $\Gamma(z)$ is the Gamma function defined by
\be
 \Gamma(z) = \int_0^\infty  t^{z-1} e^{-t}\,{\rm d}t\,.
\ee 
By replacing $z$ by $\pm i z$ one obtains the modified Bessel equation
\be
z^2 \frac{\dd^2 w}{\dd z^2} + x \frac{\dd w}{\dd z} + (z^2 + \nu^2)w = 0.
\ee
The solutions are called modified Bessel functions of first and second kind,
$I_\nu(z)$ and $K_\nu(z)$ respectively, and are connected to the functions
$J_\nu(z)$ by
\ba\label{eqn:I-J}
I_\nu(z) &=& i^{-\nu} J_\nu(iz), \\
K_\nu(x) &=& \frac{\pi}{2} \frac{I_{-\nu} (x) - I_\nu (x)}{\sin (\nu \pi)}
\ea
Unlike the ordinary Bessel functions, which are oscillating as functions of a
real argument, the modified Bessel functions are exponentially growing and
decaying functions, respectively. For real arguments $x$, the function $I_\nu(x)$ goes to zero at $x=0$ for $\nu > 0$ and is finite at $x=0$
for $\nu=0$, like the ordinary Bessel function $J_\nu(x)$. Analogously, $K_\nu(x)$ diverges at $x=0$.

There are many integral representations of these functions, using integrals
along the real axis as well as contour integrals in the complex plane.
In the following we focus on the modified Bessel functions of the second kind.
Very useful integral representations are
\ba
K_\nu(z)&=&\frac{\sqrt{\pi}(\hh z)^\nu}{\Gamma(\nu+\hh)}\int_1^\infty \!\!\!\!\!\dd t\;
{\rm e}^{-zt}(t^2-1)^{\nu-\hh}, \;\; \Re{\nu}>-\hh,\;\; \vert{\rm ph}\;
  z\vert<\hh\pi\\
K_\nu(z)&=&\int_0^\infty \!\!\!\!\!\dd t\;
{\rm e}^{-z\cosh t}\cosh(\nu t), \;\; \vert{\rm ph
  }\;z\vert<\hh\pi.
\ea
%For real arguments $x$ a useful representation for the calculus of the Feynman
%propagator in field theory is
%\be
%K_\nu(x) = \frac{1}{2} {\rm e}^{-\frac{1}{2}\nu\pi i}
%\int_{-\infty}^{+\infty}\!\!\!\!\!\dd t\;
%{\rm e}^{-ix\sinh t -\nu t}.
%\ee
Like the ordinary Bessel function, the modified Bessel function of second kind
can be expressed as a power series.
In the case of integer order the expansion reads
\ba
K_n(z)&=&\hh(\hh
z)^{-n}\sum_{k=0}^{n-1}\frac{(n-k-1)!}{k!}(-\frac{1}{4}z^2)^k+(-1)^{n+1}\ln
(\hh z)I_n(z)\nonumber\\
&+& (-1)^n\hh(\hh z)^n\sum_{k=0}^{\infty}(\psi(k+1)+\psi(n+k+1))\frac{(\frac{1}{4}z^2)^k}{k!(n+k)!}, 
\ea
where $\psi(w)=\Gamma'(w)/\Gamma(w)$ denotes the $\psi$ function, which satisfies
\be\label{eqn:Psi}
\psi(1)=-\gamma_{\rm EM} \quad\mbox{and}\quad\psi(n+1)=\sum_{k=1}^n\frac{1}{k}-\gamma_{\rm EM}.
\ee
Here $\gamma_{\rm EM}$ denotes the Euler-Mascheroni constant.
%defined as the
%limiting difference between the harmonic series and the natural logarithm
%\be
%\gamma_{\rm EM} = \lim_{n \rightarrow \infty } \left( 
%\sum_{k=1}^n \frac{1}{k} - \ln(n) \right).
%\ee
The expansion for $K_2(z)$ and $K_3(z)$ is obtained by combination of
Eq. (\ref{eqn:J-exp}) and (\ref{eqn:I-J}) together with the definitions in
Eq. (\ref{eqn:Psi}). 
In addition to the normalization-like condition 
\be
\int_0^\infty \!\!\!\!\!\dd t\;
K_\nu(t)=\hh\pi\sec (\hh\pi\nu), \quad \vert\Re{\nu}\vert<1
\ee
we recall here 
\be
\int_0^\infty \!\!\!\!\!\dd t\;
t^{\mu-1}K_\nu(t)=2^{\mu-2}\Gamma \left(\hh\mu-\hh\nu\right)\Gamma\left(\hh\mu+\hh\nu\right), \quad
\vert\Re{\nu}\vert<\Re{\mu}, 
\ee
which is equivalent to the expression in Eq. (\ref{eqn:K-int}).

\bibliographystyle{elsarticle-num-names}
\bibliography{literature}

\end{document}